\documentclass[orivec]{llncs}
\usepackage[utf8]{inputenc}
\usepackage[T1]{fontenc}
\usepackage{amsmath,stmaryrd,mathpartir}
\usepackage{amssymb}
\usepackage{proof}
\usepackage[]{graphicx}
\usepackage{paralist}
\usepackage{enumitem}
\usepackage{color}
\usepackage{xspace}
\usepackage{url}
\usepackage{bussproofs}
\usepackage{appendix}
\usepackage[inference]{semantic}
\usepackage{fancyhdr}
\usepackage{epsfig}
\usepackage{latexsym}
\usepackage{lmodern}
\usepackage{microtype}
\usepackage{lastpage} 
\usepackage{hyperref}
\usepackage{tabularx}
\usepackage{multirow}
\usepackage{listings}
\lstdefinestyle{customhaskell}{
    language=Haskell,
    showstringspaces=false,
    basicstyle=\footnotesize\rmfamily,
    keywordstyle=\bfseries\color{green!40!black},
    commentstyle=\itshape\color{blue!40!black},
    identifierstyle=\color{black},
    stringstyle=\color{orange!40!black},
}
\usepackage[dvipsnames*,svgnames]{xcolor}
\hypersetup{colorlinks=true,allcolors=DarkBlue,linkcolor=black}
\newcommand{\setof}[1]{\{ #1 \}}
\newcommand{\ap}[2]{\mbox{$\mathit{#1}(#2)$}}
\newcommand{\bap}[3]{\mbox{$\mathit{#1}(#2,#3)$}}
\newcommand{\tap}[4]{\mbox{$\mathit{#1}(#2,#3,#4)$}}

\newcommand{\step}[1]{\mathbin{\lower0.55ex\hbox{$\lhook\joinrel\xrightarrow{#1}$}}}
\newcommand{\semstep}[1]{\step{#1}}

\newcommand{\Activity}{\mathsf{Activity}}

\newcommand{\ContProv}{\mathsf{ContProv}}

\newcommand{\Comp}{\mathsf{Comp}}

\newcommand{\CompInstance}{\mathsf{CompInstance}}

\newcommand{\iComp}{\mathsf{iComp}}
\newcommand{\Val}{\mathsf{Val}}
\newcommand{\PermId}{\mathsf{PermId}}
\newcommand{\PermLvl}{\mathsf{PermLvl}}
\newcommand{\Perm}{\mathsf{Perm}}

\newcommand{\PermGrp}{\mathsf{PermGroup}}

\newcommand{\Uri}{\mathsf{Uri}}

\newcommand{\Manifest}{\mathsf{Manifest}}

\newcommand{\AppId}{\mathsf{AppId}}
\newcommand{\Cert}{\mathsf{Cert}}
\newcommand{\Res}{\mathsf{Res}}
\newcommand{\ResVal}{\mathsf{Val}}

\newcommand{\SysApp}{\mathsf{SysImgApp}}

\newcommand{\Intent}{\mathsf{Intent}}

\newcommand{\InstApps}{\mathsf{InstApps}}
\newcommand{\VerifiedApps}{\mathsf{VerifiedApps}}
\newcommand{\GrantedGroups}{\mathsf{PermsGr}}
\newcommand{\AppsPerms}{\mathsf{AppPS}}
\newcommand{\AppsDefPerms}{\mathsf{AppDefPS}}
\newcommand{\CompInsRunning}{\mathsf{CompInsRun}}
\newcommand{\OpType}{\mathsf{OpTy}}

\newcommand{\DelPPerms}{\mathsf{DelPPerms}}

\newcommand{\DelTPerms}{\mathsf{DelTPerms}}

\newcommand{\AppsResCont}{\mathsf{ARVS}}
\newcommand{\SentIntents}{\mathsf{Intents}}
\newcommand{\AppsCert}{\mathsf{Certs}}
\newcommand{\AppsManifest}{\mathsf{Manifests}}
\newcommand{\ImageApps}{\mathsf{SysImage}}
\newcommand{\AndroidState}{\mathsf{AndroidST}}

\newcommand{\Action}{\mathsf{Action}}
\newlength{\bcextramargin}
\newenvironment{changemargin}[2]{\begin{list}{}{%
\setlength{\topsep}{0pt}%
\setlength{\leftmargin}{0pt}%
\setlength{\rightmargin}{0pt}%
\setlength{\listparindent}{\parindent}%
\setlength{\itemindent}{\parindent}%
\setlength{\parsep}{0pt plus 1pt}%
\addtolength{\leftmargin}{#1}%
\addtolength{\rightmargin}{#2}%
}\item }{\end{list}} 

\newcommand{\actdefsection}[1]{
\begin{changemargin}{-\bcextramargin}{0pt}
\vspace{1ex}
\noindent
\textbf{{#1}}
\end{changemargin}
}

\newenvironment{absolutelynopagebreak}
  {\par\nobreak\vfil\penalty0\vfilneg
   \vtop\bgroup}
  {\par\xdef\tpd{\the\prevdepth}\egroup
   \prevdepth=\tpd}

\newcommand{\Mathexecrel}[3]{#1 \semstep{#2} #3}
\newtheorem{prop}{Property}
\newcommand{\eqdef}{\stackrel{{\rm def}}{=}}

\newcommand{\lemref}[1]{Lemma~\ref{#1}}
\begin{document}
\title{Towards a certified reference monitor of the Android 10 permission system
} \author{Guido De Luca\inst{1} \and Carlos
    Luna\inst{2}} \institute{Universidad Nacional de Rosario, Argentina. 
   \\ \email{gdeluca@dcc.fceia.unr.edu.ar} 
\and InCo, Facultad de Ingenier\'ia, Universidad de la Rep\'ublica, Uruguay.
 \\ \email{cluna@fing.edu.uy}
}

\maketitle

\begin{abstract}
Android is a platform for mobile devices that captures more than 85\%  of the total market-share~\cite{reporteIDC}. Currently, mobile devices allow people to develop multiple tasks in different areas. Regrettably, the benefits of using mobile devices are counteracted by increasing security risks. The important and critical role of these systems makes them a prime target for formal verification. 
In our previous work~\cite{LOPSTR:BetarteCGL}, we exhibited a formal specification of an idealized formulation of the permission model of version \texttt{6} of Android.
In this paper we present an enhanced version of the model in the proof-assistant \texttt{Coq}, including the most relevant changes concerning the permission system introduced on versions \texttt{Nougat}, \texttt{Oreo}, \texttt{Pie} and \texttt{10}.
The properties that we had proved earlier for the security model has been either revalidated or
refuted, and new ones have been formulated and proved.
Additionally, we make observations on the security of the most recent versions of Android.
Using the programming language of \texttt{Coq} we have developed a functional implementation of a reference validation mechanism and certified its correctness. 
The formal development is about 23k LOC of \texttt{Coq}, including proofs. 


\end{abstract}
\section{Introduction}
\label{sec:intro}

Android \cite{AndroidProy} is the most used mobile OS in the world, capturing approximately the 85\% of the total market-share \cite{reporteIDC}. It offers a
huge variety of applications in its official store that aim to help people in their daily activities, many of them critical in terms of privacy. In order to
guarantee their users the security they expect, Android relies on a \textit{multi-party} consensus system where user, OS and application must be all in favour
of performing a task. This security framework is built upon a system of permissions, which are basically tags that developers declare on their
applications to gain access to sensitive resources. Whenever an action that requires some of this permissions is executed for the first time, the user will be
asked for authorization and if provided, the OS will ensure that only the required access is granted. The important and critical role of this security mechanism
makes it a prime target for (formal) verification.

Security models play an important role in the design and evaluation of security mechanisms of systems. Earlier, their importance was already pointed in the Anderson report \cite{Anderson:1972}, where the concept of \emph{reference monitor} was first introduced. This concept defines the design requirements for implementing what is called a \emph{reference validation mechanism}, which shall be responsible for enforcing the access control policy of a system. For ensuring the correct working of this mechanism three design requirements are specified:
\begin{inparaenum}[i)]
\item the reference validation mechanism (RVM) must always be invoked (\emph{complete mediation});
\item the RVM must always be tamper-proof (\emph{tamper-proof}); and
\item the RVM must be small enough to be subject to analysis and tests, the completeness of which can be assured (\emph{verifiable}).
\end{inparaenum}

The work presented here is concerned with the verifiability requirement. In particular we put forward an approach where formal analysis and verification of properties is performed on an idealized model that abstracts away the specifics of any particular implementation, and yet provides a realistic setting in which to explore the issues that pertain to the realm of (critical) security mechanisms of
Android.
The formal specification of the reference monitor shall be used to establish and prove that the security properties that constitute the  intended access control policy are satisfied by the modeled behavior of the validation mechanisms.

\subsubsection*{Contributions} 
In our previous work~\cite{LOPSTR:BetarteCGL} we  presented a formal specification of an idealized formulation of the permission model of version 6 of Android. We
also developed, using the programming language of \texttt{Coq} \cite{coq-manual}, an executable (functional) specification of the reference validation mechanism
and we proved its correctness conforming the specified model. Lastly, we used the program extraction mechanism provided by \texttt{Coq} \cite{conf/types/Letouzey02} to derive a certified
\texttt{Haskell} implementation of the reference validation mechanism.
Here we present an enhanced version of the model, including the most relevant changes concerning the permission system introduced on versions \texttt{Nougat}, \texttt{Oreo}, \texttt{Pie}
and \texttt{10}. Some of these changes don't have a direct impact on our abstract model. In those cases, an informal analysis is included. The executable specification
was also updated, and with that, the derived implementation as well. The properties that we had proved for the security model has been either revalidated or
refuted, and new ones have been formulated and proved. The formal development is about 23k LOC of \texttt{Coq}, including various lemmas and their proofs.

\subsubsection*{Organization of the paper}
Section~\ref{sec:background} reviews the security mechanisms of Android and briefly describes the changes introduced in the later versions.
Sections~\ref{sec:model}~and~\ref{sec:excspec} present the formal axiomatic specification and the semantics of the certified implementation, respectively.
Both sections discuss relevant properties concerning the new features. Section~\ref{sec:relwork} considers related work and finally,
Section~\ref{sec:conclusion} concludes with a summary of our contributions and directions for future work. The full formalization is available at: \\ \url{https://www.fing.edu.uy/~cluna/Android10-Coq.zip} \cite{AndroidCoq:2020} and can be verified using the \texttt{Coq} proof assistant. 


\section{Android's security model}
\label{sec:background}
\subsection{Basic security mechanisms}
  The Android security model is primarily based on a sandbox and permission mechanism. Each application runs in a private virtual machine with a unique ID
  assigned to it, which means that one application's code is isolated from the code of all others. This way of protection entails the existence of a decision
  procedure (a reference validation mechanism) that arbitrates the access to sensitive data whenever an application wants to share resources with another (or
  the system). Decisions are made by following security policies using a simple notion of permission. 
  
  Every permission is identified by a unique name/text, has a protection level and may belong to a permission group. Furthermore, permissions can be classified
  into two groups: the ones defined by an application, for the sake of self-protection; and those predefined by Android, which are required to gain access to
  certain system features, like internet or location. 
  Depending on the protection level of the permission, the system defines the corresponding decision procedure. There are three classes of permission
  levels~\cite{protectionLevel}:
  \begin{inparaenum}[i)] 
    \item \textit{normal}, this permissions can be automatically granted since they cover data or resources where there’s very little risk to the user’s
    privacy or the operation of other apps;
    \item \textit{dangerous}, permissions of this level provide access to data or resources that may be sensitive or could potentially affect the operation of
    other applications, and explicit user approval is needed to be granted; and
    \item \textit{signature}, a permission of this level is granted only if the application that requires it and the application that defined it are both signed
    with the same certificate.
  \end{inparaenum}
  An application must declare --in an XML file called \texttt{AndroidManifest}-- the set of permissions it needs to acquire further capacities than the
  default ones. From version \texttt{6} of Android, \textit{dangerous} permissions are granted on runtime whereas both \textit{normal} and \textit{signature} are given
  when the application is installed.

  Permissions may belong to groups that reunites a device's capabilities. The main purpose of grouping permissions this way is
  to handle permission requests at the group level, in order to avoid overwhelming the user with too many questions. For
  example, the \texttt{SMS} group includes the permission needed to read the text messages as well as the one needed to receive
  them (both considered to be \textit{dangerous}). Whenever an application needs one of those for the first time, the user will
  be asked to authorize the whole group. On Section~\ref{sec:background:changes:oreo}, we explain what \textit{authorizing
  a group} means depending on the platform version.
 

  An Android application is built up from \textit{components}. A component is a basic unit that provides a particular functionality and that can be run by any
  other application with the right permissions. There exist four types of components \cite{fundamentals}: 
  \begin{inparaenum}[i)]
    \item \textit{activity}, which is essentially a user interface of the application; 
    \item \textit{service},  a component that executes in background without providing an interface to the user;
    \item \textit{content provider},  a component intended to share information among applications; and 
    \item \textit{broadcast receiver}, a component whose objective is to receive messages, sent either by the system or an application, and trigger the
    corresponding actions. 
  \end{inparaenum}
  The communication between components is achieved with the exchange of special messages called \textit{intents}, which can be either 
  \begin{inparaenum}[i)]
    \item \textit{explicit}, meaning that the target application is specified; or
    \item \textit{implicit}, where only the action to be performed is declared and the system determines which application will run the task (if
    there is more than one capable application, the user is allowed to choose). 
  \end{inparaenum}
  In order to be candidates for the implicit intents resolutions, an application must declare on their manifest an \textit{intent filter} that indicates the
  types of intents it can respond to.

  Android provides two mechanisms by which an application can delegate its own permissions to another one. These mechanisms are called \textit{pending intents}
  and \textit{URI permissions}. An intent may be defined by a developer to perform a particular action. A \texttt{PendingIntent} specifies a reference to an
  action, which might be used by another application to perform the operation with the same permissions and identity of the one that created the intent. The
  \textit{URI permissions} mechanism can be used by an application that has read/write access to a \textit{content provider} to temporarily delegate those
  permissions to another application. These permissions are revoked once the receiver activity or service become inactive.

\subsection{A brief review on the changelog}
\label{sec:background:changes}
As we described in our previous work~\cite{LOPSTR:BetarteCGL}, the sixth version of Android introduced an important change to the system, allowing the users to
handle permissions at runtime. In this section, we give a short account of the changes introduced between Android \texttt{Nougat} (\texttt{7}) and Android \texttt{10}, that had a
significant impact on the permission system.

\subsubsection*{Filesystem}
In order to improve security, the private directory of applications targeting\footnote{Applications can \textit{target} a particular version of the system.
Android uses this setting to determine whether to enable any compatibility behaviors or features.} Android \texttt{7.0} or higher has restricted access: only the owner
is capable of reading, writing or executing files stored in there. This configuration prevents leakage of metadata of private files, like the size or existence.
With this change, applications are no longer able to share files simply by changing the file permissions and sharing their private URI; a content provider must
be used in order to generate a reference to the file. With this approach, a new kind of URI is generated, which grants a temporary permission that will be
available for the receiver activity or service only while they are active/running.

Our previous model already allowed granting temporary permissions to content providers URIs, so no change was required to formalize this new feature.

\subsubsection*{Grouped permissions}
\label{sec:background:changes:oreo}
Prior to Android \texttt{8}, if an application requested a grouped permission at runtime and the user authorized it, the system also granted the rest of the permissions
from the same group that were declared on the manifest. This behaviour was incorrect since it violated the intended least privilege security policy claimed by
the designers of the platform. For applications targeting Android \texttt{8} or higher, this action was corrected and only the requested permission is granted. However,
once the user authorized a group, all subsequent requests for permissions in that group are automatically granted.
This change was added to the model. 

\paragraph*{Normal grouped permissions.} According to Android's official documentation, \textit{any permission can belong to a permission group regardless of
protection level}~\cite{permissions}. However, it is not specified if normal and dangerous permissions can share a group nor, in case that it is possible, how
the system should handle this situation. A few questions we have raised ourselves are the following: 
\begin{inparaenum}[i)]
  \item Is the authorization to automatically concede permissions from that group granted at installation time together with the normal permissions?;
  \item Is the user warned about that decision?;
  \item If that is the case, then there's a contradiction with the documentation, since it claims that \textit{a permission's group only affects the user
  experience if the permission is dangerous}; and 
  \item If it's not, how does the system avoid that dangerous permissions from the same group are not automatically granted later by the system?
\end{inparaenum}

In this work we formalized a worst-case scenario (that still suits the informal specification given by the authors of the platform), where a normal permission
enables the automatic granting of dangerous permissions belonging to the same group. We formally discuss this situation in Section~\ref{sec:properties}.


\subsubsection*{Privacy changes}
Android \texttt{Pie} (\texttt{9}) introduced several changes aiming to enhance users' privacy, such as limiting background apps' access to device sensors, restricting information
retrieved from Wi-Fi scans, and adding new permission groups and rules to reorganize phone calls and phone state related permissions. Later, the tenth version
of the platform continued adding limitations to services: a new permission for accessing the location in the background was added. Furthermore, Android \texttt{10}
placed restrictions on when a service can start an activity, in order to minimize interruptions for the user and keep the user more in control of what's shown
on their screen.

These changes are specific to the implementation, meaning that they have no impact on an abstract representation like ours.

\subsubsection*{Permission check on legacy apps}
\label{sec:background:changes:10}
Applications that target Android \texttt{5.1} or lower are considered to be old\footnote{We can also refer to them as
\textit{legacy} applications.}. If an \textit{old} application runs on an Android \texttt{10} system for the first time, a
prompt appears on the screen, giving the user an opportunity to revoke access to permissions that the system previously granted
at install time. This feature has been added to our model.

\section{Formalization of Android's permission system}
\label{sec:model}
In this section we describe the axiomatic semantics of our model of the system, focusing on the features introduced in
the later versions. We also discuss some of the verified properties.

\subsubsection*{Formal language used} 
\texttt{Coq} is an interactive theorem prover based on higher order logic that allows to write formal specifications
and interactively generate machine-checked proofs of theorems. It also provides a (dependently typed) functional programming
language that can be used to write executable algorithms. The \texttt{Coq} environment also provides program extraction towards
languages like Ocaml and Haskell for execution of (certified) algorithms \cite{letouzey04,conf/types/Letouzey02}.
In this work, enumerated types and sum types are defined using Haskell-like notation; for example, $option\ T \eqdef None \mid
Some\ (t:T)$. Record types are of the form $\left\{ l_1 : T_1,\ldots, l_n : T_n \right\}$, whereas their elements are of the
form $\{t_1, \ldots, t_n \}$. Field selection is written as $r.l_i$. We also use $\setof{T}$ to denote the set of elements of
type $T$. Finally, the symbol $\times$ defines tuples, and $nat$ is the datatype of natural numbers. We omit \texttt{Coq} code
for reasons of clarity; this code is available in \cite{AndroidCoq:2020}.

\subsection{Model states}
\label{sec:model:stateI}

The Android security model we have developed has been formalized as an abstract state machine. In this model, states ($\AndroidState$) are modelled as
13-tuples that store dynamic data about the system such as the installed applications and their current permissions, as well as static data like the
declared manifest of each installed app. A complete formal definition is given in  Figure~\ref{fig:model:state}.


\begin{figure}
  \input{state_figure}
\caption{Android state}
\label{fig:model:state}
\end{figure}

The type $\PermId$ represents the set of permissions identifiers;
$\PermGrp$, the set of permission groups identifiers;
$\Comp$, the application components whose code will run on the system;
$\AppId$ represents the set of application identifiers;
$\iComp$ is the set of identifiers of running instances of application components;
$\ContProv$ is a subset of $\Comp$, a special type of component that allows sharing resources among different applications;
a member of the type $\Uri$ is a particular URI (uniform resources identifier);
the type $\Res$ represents the set of resources an application can have (through its $content~providers$, members of $\ContProv$);
the type $\Val$ is the set of possible values that can be written on resources;
an intent --i.e. a member of type $\Intent$-- represents the intention of a running component instance to start or communicate with other applications;
a member of $\SysApp$ is a special kind of applications which are deployed along with the OS itself and has certain privileges, like being impossible to uninstall. 

The first component of the state records the identifiers ($\AppId$) of the applications installed by the user. The second component is a subset of the first
one, and represents those applications that are considered to be old but have already been verified, also by the user. 
The third component keeps track of the permissions granted to every application present in the system, including the ones preinstalled on the system. Similarly,
the next component holds the information about what permission groups have already been authorized by the user on each app.
The fifth component of the state stores the set of running component instances ($\CompInstance$), while the components $\DelPPerms$ and $\DelTPerms$ store the
information concerning permanent and temporary permissions delegations, respectively\footnote{A permanent delegated permission represents that an app has
delegated permission to perform an operation on the resource identified by an URI. A temporary delegated permission refers to a permission that has been
delegated to a component instance.}.
The eighth and ninth components of the state store respectively the values ($\Val$) of resources ($\Res$) of applications and the set of intents ($\Intent$) sent
by running instances of components ($\iComp$) not yet processed. The four last components of the state record information that represents the manifests of the
applications installed by the user, the certificates ($\Cert$) with which they were signed and the set of permissions they define. The last component of the
state stores the set of (native) applications installed in the Android system image, information that is relevant when granting permissions of level
$Signature/System$.

A manifest ($\Manifest$) is modelled as a 6-tuple that respectively declare application components (set of components, of type $\Comp$, included in the application); optionally, the minimum version of the Android SDK required to run the application; optionally,  the version of the Android SDK targeted on development; the set of permissions it may need to run at its maximum capability; the set of permissions it declares; and the permission required to interact with its components, if any. 
Application components are all denoted by a component identifier. 
A content provider ($\ContProv$),  in addition, encompasses a mapping to the managed resources from the URIs assigned to them for external access. 
While the components constitute the static building blocks of an application, all runtime operations are initiated by component instances, which are represented in our model as members of an abstract type. 

\subsubsection*{Valid states}
\label{sec:model:executions:valid_states}
The states defined this way include some cases that are not relevant with the model we are trying to analyze. For example, we
don't want a state were a preinstalled application and one installed by the user have the same identifier. In order to prevent
this inconsistencies, we define a notion of valid state that captures several well-formedness conditions. It is formally defined
as a predicate \texttt{valid\_state} on the elements of type $\AndroidState$. This predicate holds on a state $s$ if the
following conditions are met: 
\begin{itemize} 
\item all the components both in installed applications and in system applications have different identifiers;
\item no component belongs to two different applications present in the device;
\item no running component is an instance of a content provider;
\item every temporally delegated permission has been granted to a currently running component and over a content provider present in the system;
\item every running component belongs to an application present in the system;
\item every application that sets a value for a resource is present in the system;
\item the domains of the partial functions $\AppsManifest$, $\AppsCert$ and $\AppsDefPerms$ are exactly the identifiers of the user installed applications;
\item the domains of the partial functions $\AppsPerms$ and $\GrantedGroups$ are exactly the identifiers of the applications in the system, both those installed by the users and the system applications;
\item every installed application has an identifier different to those of the system applications, whose identifiers differ as well;
\item all the permissions defined by applications have different identifiers;
\item every partial function is indeed a function, that is, their domains don't have repeated elements;
\item every individually granted permission is present in the system; and
\item all the sent intents have different identifiers.
\end{itemize}
All these safety properties have a straightforward interpretation in our model. The full formal definition of the predicate is
available in \cite{AndroidCoq:2020}.


\subsection{Action semantics}
\label{sec:model:actsemI}
We modelled the different functionalities provided by the Android security system as a set of actions (of type $\Action$) that determine how the system is able
to transition from one state to another.  Table \ref{table:old_actions} summarises the actions specified in our previous model that remained mostly the same
since the new features didn't affect them whereas Table \ref{table:new_actions} groups those that are new or that suffered a big semantic change.


\begin{table}[thb!]
\scriptsize
\centering
\begin{tabularx}{\linewidth}{|l X|}
	\hline
	$\mathtt{install}~app~m~c~lRes$	& Install application with id $app$, whose manifest is $m$, is signed with certificate $c$ and its resources list is $lRes$. \\
	\hline
	$\mathtt{uninstall}~app$	& Uninstall the application with id $app$. \\
	\hline
	$\mathtt{read}~ic~cp~u$	& The running comp. $ic$ reads the resource corresponding to URI $u$ from content provider $cp$. \\
	\hline
	$\mathtt{write}~ic~cp~u~val$	& The running comp. $ic$ writes value $val$ on the resource corresponding to URI $u$ from content provider $cp$. \\
	\hline
	$\mathtt{startActivity}~i~ic$	& The running comp. $ic$ asks to start an activity specified by the intent $i$. \\
	\hline
	$\mathtt{startActivityRes}~i~n~ic$	& The running comp. $ic$ asks to start an activity specified by the intent $i$, and expects as return a token $n$. \\
	\hline
	$\mathtt{startService}~i~ic$	& The running comp. $ic$ asks to start a service specified by the intent $i$. \\
	\hline
	$\mathtt{sendBroadcast}~i~ic~p$	& The running comp. $ic$ sends the intent $i$ as broadcast, specifying that only those components who have the permission $p$ can receive it. \\
	\hline
	$\mathtt{sendOrdBroadcast}~i~ic~p$	& The running comp. $ic$ sends the intent $i$ as an ordered broadcast, specifying that only those components who have the permission $p$ can receive it. \\
	\hline
	$\mathtt{sendSBroadcast}~i~ic$	& The running comp. $ic$ sends the intent $i$ as a sticky broadcast. \\
	\hline
	$\mathtt{resolveIntent}~i~app$	& Application $app$ makes the intent $i$ explicit. \\
	\hline
	$\mathtt{stop}~ic$	& The running comp. $ic$ finishes its execution. \\
	\hline
	$\mathtt{grantP}~ic~cp~app~u~pt$	& The running comp. $ic$ delegates permanent permissions to application $app$. This delegation enables $app$ to perform operation $pt$ on the resource assigned to URI $u$ from content provider $cp$. \\
	\hline
	$\mathtt{revokeDel}~ic~cp~u~pt$	& The running comp. $ic$ revokes delegated permissions on URI $u$ from content provider $cp$ to perform operation $pt$. \\
	\hline
	$\mathtt{call}~ic~sac$	& The running comp. $ic$ makes the API call $sac$. \\
	\hline
  \end{tabularx}
  \caption{Legacy actions}
  \label{table:old_actions}
\end{table}

\begin{table}[thb!]
\scriptsize
\centering
\begin{tabularx}{\linewidth}{|l X|}
	\hline
	$\mathtt{grant}~p~app$	& Grant the permission $p$ to the application $app$ with user confirmation. \\ 
	\hline
	$\mathtt{grantAuto}~p~app$	& Grant automatically the permission $p$ to the application $app$ (without user confirmation). \\
	\hline
	$\mathtt{revoke}~p~app$	& Remove an ungrouped permission $p$ from the application $app$. \\
	\hline
	$\mathtt{revokePermGroup}~g~app$	& Remove the every permission of group $g$ from the application $app$. \\
	\hline
	$\mathtt{hasPermission}~p~app$	& Check if the application $app$ has the permission $p$. \\
	\hline
	$\mathtt{receiveIntent}~i~ic~app$	& Application $app$ receives the intent $i$, sent by the running comp. $ic$. \\
	\hline
	$\mathtt{verifyOldApp}~app$	& Application $app$ granted permissions are verified by the user  \\
	\hline
  \end{tabularx}
  \caption{New or modified actions}
  \label{table:new_actions}
\end{table}

The behaviour of each action is specified in terms of a precondition ($ Pre  :  \AndroidState \rightarrow \Action \rightarrow Prop $) and a postcondition ($
Post  :  \AndroidState \rightarrow \Action \rightarrow \AndroidState \rightarrow Prop $). For instance, the axiomatic semantics of the new feature about
automatic granting of permissions \texttt{grantAuto} is given by:
\small
$$\begin{array}{l}
\quad \quad Pre(s,\texttt{grantAuto}\ p\ app) \eqdef\\ 
\quad \quad \quad \quad (\exists m:\Manifest,\ m = getManifestForApp(app,s) \\
\quad \quad \quad \quad \quad \quad \land\ getPermissionId(p) \in (use\ m))\ \land\ \\
\quad \quad \quad \quad (isSystemPerm\ p\ \lor usrDefPerm\ p)\ \land\ \\
\quad \quad \quad \quad p\ \notin grantedPerms(app,s)\ \land\ \\
\quad \quad \quad \quad permLevel(p)\ = dangerous\ \land\ \\
\quad \quad \quad \quad (\exists g:\PermGrp,\  getPermissionGroup(p) = Some\ g\ \\
\quad \quad \quad \quad \quad \quad  \land\ g \in getAuthorizedGroups(app,s))  \\

\quad \quad Post(s,\texttt{grantAuto}\ p\ app, s') \eqdef\\ 
\quad \quad \quad \quad grantPerm(app,p,s,s')\ \land\ \\
\quad \quad \quad \quad sameOtherFieldsOnGrantAuto(s,s')\
\end{array}$$ 

\normalsize 

The precondition establishes several conditions that must be fulfilled before this action is able to transition. The first one requires that the permission $p$
is listed on the application's manifest (and this manifest, of course, is required to exist). Regarding the permission, it's also required that its defined
either by the user or the system, that its level is $dangerous$ and that it has not been already granted to $app$. Up to this point, the precondition of
$grantAuto$ is exactly the same than the one of $grant$. The main difference is established by the following condition: the permission at issue should belong to
a group $g$ and the system should know that the user had previously authorized that group for automatic granting.

The postcondition of $\texttt{grantAuto}\ p\ a$ requires that for an initial state $s$ and a final state $s'$, the individual
permission $p$ is granted to application $app$. This condition is enforced by the $\texttt{grantPerm}\ a\ p\ s\ s'$ predicate which only alters the state in
component that maps applications with their current \textit{dangerous} permissions. Every other component of the state remains the same.

\subsection{Executions}


\label{sec:model:executions}
Whenever the system attempts to execute an action $a$ over a valid state $s$, there are two possible outcomes. If the precondition holds, the system will
transition to another state $s'$ where the postcondition of $a$ is established; but if the precondition is not satisfied on $s$, then the state remains
unchanged and the system answers with an error message determined by the relation $ErrorMsg$\footnote{Given a state $s$, an action $a$ and an error code $ec$,
$\tap{ErrorMsg}{s}{a}{ec}$ holds iff $error~ec$ is an acceptable response when the execution of $a$ is requested on state $s$.}.

Formally, the possible answers of the system are defined by the type $$\textit{Response} \eqdef ok\ |\ error\ (ec : ErrorCode)$$ and the executions can be
specified with this operational semantics:
\footnotesize
\begin{displaymath}
\begin{array}{c}
\inference[]{$$valid\_state(s)$$ \hspace{.2cm} $$Pre(s, a)$$ \hspace{.2cm} $$Post(s, a, s')$$}{$$s\step{a/ok}s'$$} 
\hspace{0.5cm} 
\inference[]{$$valid\_state(s)$$ \hspace{.2cm} $$ErrorMsg(s, a, ec)$$}{$$s\step{a/error(ec)}s$$}
\end{array}
\end{displaymath}
\normalsize

One-step execution with error management preserves valid states.
\begin{lemma} [Validity is invariant]
\footnotesize
\label{lemma:valid-state-correct}
\mbox{}\\
$\begin{array}{l}
\forall\ (s\ s':\AndroidState)(a:\Action) (r:Response),
s\step{a/r}s' \rightarrow valid\_state(s')
\end{array}$
\end{lemma}
The property is proved by case analysis on $a$, for each condition in \textit{valid\_state},
using several auxiliary lemmas \cite{AndroidCoq:2020}.

System state invariants, such as state validity, are useful to analyze other relevant properties of the model.  In particular, 
the results presented in this work are obtained from valid states of the system.

\subsection{Reasoning over the specified model}
\label{sec:properties}


In this section we present and discuss some properties about the Android \texttt{10} security framework.  We focus on safety-related properties about the changes
introduced on the later versions of Android (mainly \texttt{Oreo} and \texttt{10}) rather than on security issues. Nevertheless, we also found potentially
dangerous behaviours that may not be considered in the informal documentation of the platform and we formally reasoned about them as well.  The full formal
definition of these properties can be found in \cite{AndroidCoq:2020}, along with the corresponding proofs.

On Table \ref{tab:model:semantic_descriptions} we introduce helper functions and predicates used to define the properties that will follow.

\begin{table}[thb!]
\scriptsize
\centering
\begin{tabularx}{\linewidth}{|l X|}
\hline
\textbf{Function/Predicate} & \textbf{Description} \\
\hline
$appHasPermission(app,p,s)$ & holds iff $app$ is considered to have permission $p$ on state $s$. \\
\hline
$canGrant(cp,u,s)$	& holds iff the content provider $cp$ allows the delegation of permissions over the resource at URI $u$ on state $s$. \\
\hline
$\tap{\mathit{canStart}}{c'}{c}{s}$ & holds if the app containing component $c'$ (installed in $s$) has the required permissions to create a new running instance of $c$. \\
\hline
$cmpProtectedByPerm(c)$	& returns the permission by which the component $c$ is protected. \\
\hline
$componentIsExported(c)$	& holds iff the component $c$ is exported and can be accessed from other applications. \\
\hline
$existsRes(cp,u,s)$	& holds iff the URI $u$ belongs to the content provider $cp$ on $s$. \\
\hline
$getAppFromCmp(c,s)$	& given a component $c$ on $s$, returns the app to which it belongs. \\
\hline
$getAppRequestedPerms(m)$ & given the manifest $m$ of an app, returns the set of permissions it uses. \\
\hline
$getDefPermsApp(app,s)$	& returns the set of permissions defined by $app$ on state $s$. \\
\hline
$getGrantedPermsApp(app,s)$	& returns the set of indvidual permissions granted to $app$ on $s$. \\
\hline
$getAuthorizedGroups(app,s)$ & returns the set of permission groups that have been authorized for automatic granting for $app$ on $s$. \\ 
\hline
$getInstalledApps(s)$ & returns the set of identifiers of the applications installed on $s$.\\
\hline
$getManifestForApp(app,s)$	& returns the manifest of application $app$ on state $s$. \\
\hline
$getPermissionId(p)$	& returns the identifier of permission $p$. \\
\hline
$getPermissionLevel(p)$	& returns the permission level of permission $p$. \\
\hline
$getPermissionGroup(p)$	& returns $Some~g$ if the permission $p$ is grouped or $None$ if not. \\
\hline
$getRunningComponents(s)$	& returns the set of pairs consisting of a running instance id, and its associated component currently running on state $s$. \\
\hline
$inApp(c,app,s)$	& holds iff the component $c$ belongs to application $app$ on state $s$. \\
\hline
$permissionRequiredRead(c)$	& returns the permission required for reading the component. \\
\hline
$permSACs(p,sac)$	& holds iff permission $p$ is required for performing the system call $sac$ (of type $SACall$). \\
\hline
$oldAppNotVerified(a,s)$ & holds iff the application $a$ is considered old and the user hasn't verified it in state $s$. \\ 
\hline
\end{tabularx}
\caption{Helper functions and predicates}
\label{tab:model:semantic_descriptions}
\end{table}



The first property that we proved establishes a safety condition about the automatic granting of grouped permissions. It states
that the system is not able to transition with this action unless the group of the permission involved is already authorized.

\begin{prop} [Automatic grant only possible on authorized groups] \label{modelproperty1}
 \mbox{} \\
 \small
$\forall (s,s': \AndroidState) (p:\Perm) (g:\PermGrp) (app:\AppId),$ \\
$getPermissionLevel(p) = dangerous \land getPermissionGroup(p) = Some\ g \land  \\
g \notin getAuthorizedGroups(app)
\rightarrow\ \neg \Mathexecrel{s}{\texttt{grantAuto}~p~app/ok}{s'}$ \\ \\
\textit{Android's permission system ensures that an automatic granting can only occur on permissions that belong to authorized groups.}
\end{prop}

However, a few questions arise when trying to formally describe the situations in which \textit{a group is authorized}. The
following property formalizes an unclear behaviour: there's at least one valid state where the system can automatically grant a
grouped permission to an app even though that the application has no other permission of the same group granted at that moment.
This means that an application can have a group authorized for automatic granting via a permission that no longer exist. This is
not necessarily a security flaw. It may be a design principle to avoid asking the user to authorize the same group too many
times, but the situation is not clear nor disambiguated in the official documentation.

\begin{prop} [Auto-granting permission] \label{modelproperty2}
 \mbox{} \\
 \small
$\exists (s:\AndroidState) (p:\Perm) (g:\PermGrp) (app:\AppId),$ 
$valid\_state(s) \land \\
getPermissionLevel(p) = dangerous~ \land
getPermissionGroup(p) = Some\ g \land\\$
$\neg (\exists (p': \Perm), p' \in getGrantedPermsApp(app,s)~ \land$ \\ 
$ getPermissionGroup(p') = Some\ g) \land Pre(s, grantAuto~p~a) \\ \\$
\textit{System can automatically grant a permission even though there is currently no other permission of that group granted to the app.}
\end{prop}

The next property formalizes the situation described in Section~\ref{sec:background:changes:oreo} about normal and dangerous
permissions sharing a group. Once again, the lack of a strict documentation about the platform can conduce to
ambiguous cases or even to accept a dangerous behaviour as this one.

\begin{prop} [Dangerous permission automatically granted] \label{modelproperty3}
 \mbox{} \\
 \small
$\forall (s,s': \AndroidState)~(a: \AppId)~(m: \Manifest)~(c: \Cert)~(resources: list~\Res)~(g:\PermGrp)\\
(pDang~pNorm: \Perm), \Mathexecrel{s}{\texttt{install}~a~m~c~resources/ok}{s'} \rightarrow$ \\
$getPermissionLevel(pDang) = dangerous~ \rightarrow \\
getPermissionGroup(pDang) = Some\ g \rightarrow \\
getPermissionLevel(pNorm) = normal~ \rightarrow \\
getPermissionGroup(pNorm) = Some\ g \rightarrow \\
\{pDang,~pNorm\} \subseteq getAppRequestedPerms(m) \rightarrow \\
Pre(s', grantAuto~pDang~a)$ \\ \\
\textit{An application that uses a normal and a dangerous permission of the same group, can obtain the dangerous one
automatically after being installed.}
\end{prop}

Users are able to revoke permissions at runtime. However, the UI does not allow to revoke grouped permissions individually, the complete group should be
invalidated. Thus, we proved that whenever a user revokes a permission group from an application, every individual permission that belongs to that group is
revoked.

\begin{prop} [Revoking group revokes grouped individual permissions] \label{modelproperty4}
 \mbox{} \\
 \small
$\forall (s,s': \AndroidState)~(g:\PermGrp)~(app:\AppId),$ \\
$valid\_state(s) \rightarrow 
\Mathexecrel{s}{\texttt{revokePermGroup}~g~app/ok}{s'} \rightarrow$ \\
$\neg (\exists (p: \Perm), p \in getGrantedPermsApp(app,s')~ \land getPermissionGroup(p) = Some\ g) \\ \\$
\textit{System can automatically grant a permission even though there is currently no other permission of that group granted to the app.}
\end{prop}

The following property reasons about another change mentioned in Section \ref{sec:background:changes:10}. It reasons about a
good behaviour about the unverified legacy applications. 

\begin{prop} [Unverified old app cannot receive intents] \label{modelproperty5}
 \mbox{} \\
 \small
$\forall (s,s': \AndroidState)~(i:\Intent)~(ic:\iComp)~(app:\AppId),$ \\
$valid\_state(s) \rightarrow 
oldAppNotVerified(app, s) \rightarrow$ \\
$\neg \Mathexecrel{s}{\texttt{receiveIntent}~i~ic~app/ok}{s'}$ \\ \\
\textit{An old application that hasn't been verified by the user yet cannot receive intents, meaning that it can't start activities as well.}
\end{prop}

Finally, we include here a property that holds since version \texttt{6} of Android. Any application that wants to send information through the network must have the permission
$\mathsf{INTERNET}$, but since this permission is of level $normal$, the application just needs to declare it as used in its manifest. This will give access to
the network in an implicit and irrevocable way. Once again, this has been criticized due to the potential information leakage it allows. The following property
formally generalizes this situation and embodies a reasonable argument to roll back this security issue introduced in Android \texttt{Marshmallow}. 

\begin{prop}[Internet access implicitly and irrevocably allowed] \label{modelproperty6}
 \mbox{} \\
 \small
$ \forall (s:\AndroidState) (sac:SACall) (c:\Comp) (ic:\iComp) (p:\Perm), \\
valid\_state(s) \rightarrow permSAC(p, sac) \rightarrow \\
getPermissionLevel(p) = normal \rightarrow getPermissionId(p) \in \\
getAppRequestedPerms(getManifestForApp(getAppFromCmp(c,s),s)) \rightarrow\\
(ic, c) \in getRunningComponents(s) \rightarrow \Mathexecrel{s}{\texttt{call}~ic~sac/ok}{s}$ \\ \\
\textit{If the execution of an Android API call only requires permissions of level $normal$, it is enough for an application to list them as used on its manifest file to be allowed to perform such call.}
\end{prop}
\section{A certified reference validation mechanism}
\label{sec:excspec}


The implementation we developed in our previous model consisted in a set of \texttt{Coq} functions such that for every action in our axiomatic
specification there exists a function which stands for it. In this work we kept this approach, updating those functions for which its axiomatic counterpart
changed and adding new ones for the new actions \texttt{verifyOldApp} and \texttt{grantAuto}. 


Functions that implement actions are basically state transformers. Its definition follows this pattern: first, it is checked whether the precondition of the
action is satisfied in state $s$, and then, if that is the case, another function is called to achieve a state $s'$ where the postcondition of the action holds.
Otherwise, the state $s$ is returned unchanged along with an appropriate response specifying an error code which describes the failure. More formally, the
returned value has type $Result \eqdef \{ resp\ :\ Response, st\ :\ \AndroidState \}$. In Figure~\ref{fig:install_action} we present, as an example, the
function that implements the execution of the $\texttt{grant}$ action. The \texttt{Coq} code of this function, together with that of the remaining ones, can be
found in \cite{AndroidCoq:2020}\footnote{We omit here the formal definition of these functions due to space constraints.}. The function $grant\_pre$ is defined
as the nested validation of each of the properties of the precondition, specifying which error to throw when one of them doesn't hold. In general, every
$\texttt{<action>}\_pre$ function is defined this way. The function $grant\_post$ implements the expected behaviour of the $grant$ action: the permission
\texttt{perm} is prepended to the list\footnote{We implement the sets in the model with lists of \texttt{Coq}.} of given permissions of the application
\texttt{app} and, if that permission is grouped, that grouped is also added to the list of permissions groups authorized by the user on that application.

\begin{figure}[ht]
\footnotesize
\begin{displaymath}
\begin{array}{l}
\textbf{Definition}\ grant\_safe(perm, app, s)\ : Result\ :=  \\
\quad \textbf{match}\ grant\_pre(perm, app, s)\ \textbf{with} \\
\quad \hspace{1cm} |\ Some\ ec\ \Rightarrow\ \{\ap{error}{ec},s\}  \\
\quad \hspace{1cm} |\ None\ \Rightarrow\ \{ok, grant\_post(perm, app, s) \}\\
\quad\ \textbf{end}.\\
\end{array}
\end{displaymath}
\caption{The function that implements the \texttt{grant} action}
\label{fig:install_action}
\end{figure}

\subsubsection*{Step}
All of these functions are grouped into a \textit{step} function, which basically acts as an action dispatcher\footnote{Mechanism to trigger actions, on a
state, according to the type of event considered.}. Figure~\ref{fig:step} show the structure of this function.
\begin{figure}[ht]
\footnotesize
\begin{displaymath}
\begin{array}{l}
\textbf{Definition}\ step(s,a)\ :=\ \\
\quad \textbf{match}\ a\ \textbf{with} \\
\quad  \hspace{1cm} |\ \ldots \Rightarrow\ \ldots \\
\quad  \hspace{1cm} |\  \texttt{grant}\ perm\ app\ \Rightarrow\ grant\_safe(perm, app, s)\ : Result\ :=  \\
\quad  \hspace{1cm} |\ \ldots \Rightarrow\ \ldots \\
\quad\ \textbf{end}.
\end{array}
\end{displaymath}
\caption{Structure of the \texttt{step} function}
\label{fig:step}
\end{figure}

\subsubsection*{Traces}
\label{sec:excspec:traces}

We have modeled the execution of the permission validation mechanism during a session of the system as a function that implements the execution of a list of
actions starting in an initial system state. The output of the execution, a trace, is the corresponding sequence of states.

\footnotesize
\begin{displaymath}
\begin{array}{l}
	\textbf{Function}\ trace\ (s:\AndroidState)\ (actions:list\ \Action)\ :\ list\ \AndroidState \ := \\
	\quad \textbf{match}\ actions\ \textbf{with} \\
	\quad \hspace{1cm} |\ nil\ \Rightarrow\ nil \\
	\quad \hspace{1cm} |\ action::rest\ \Rightarrow\ \textbf{let}\ s'\ :=\ (step\ s\ action).st\ \textbf{in}\ s'::trace\ s'\ rest \\
	\quad \textbf{end.}
\end{array}
\end{displaymath}
\normalsize

\subsection{Correctness of the implementation}
\label{sec:soundness}
We proceed now to outline the proof that our functional implementation of the security mechanisms of Android correctly implements the axiomatic model. This
property has been formally stated as the following correctness theorem which in turn was verified using \texttt{Coq} \cite{AndroidCoq:2020}.

\begin{theorem} 
[Correctness of the reference validation mechanism] \label{theorem:soundness}
\footnotesize
 \mbox{} \\
$ \begin{array}[t]{l}
\forall\ (s:\AndroidState)\ (a:\Action),
\ap{valid\_state}{s} \rightarrow\ 
s\step{a/step(s,a).resp} \bap{step}{s}{a}.st \\
 \end{array} $
\end{theorem}

The proof of this theorem starts by performing a case analysis on the (decidable) predicate $Pre(s,a)$. Then, in case that the predicate holds, we apply
Lemma~\ref{lemma:soundvalid}; otherwise we continue by applying Lemma~\ref{lemma:sounderror}.


\begin{lemma} 
[Correctness of valid execution] \label{lemma:soundvalid}
 \mbox{} \\
\footnotesize
$ \begin{array}[t]{l}
\forall\ (s:\AndroidState)\ (a:\Action),
\ap{valid\_state}{s} \rightarrow\ 
Pre(s, a)\ \rightarrow\\
s \step{a/ok} \bap{step}{s}{a}.st\ \wedge\  \bap{step}{s}{a}.resp=ok\\
 \end{array} $
\end{lemma}

\begin{lemma} 
[Correctness of error execution] \label{lemma:sounderror}
 \mbox{} \\
\footnotesize
$ \begin{array}[t]{l}
\forall\ (s:\AndroidState)\ (a:\Action), 
\ap{valid\_state}{s} \rightarrow\ 
\neg Pre(s,a) \rightarrow\ \exists\ (ec:ErrorCode),\\
\bap{step}{s}{a}.st = s \wedge \bap{step}{s}{a}.resp = error(ec) \wedge  ErrorMsg(s,a,ec)\\
 \end{array} $
\end{lemma}

	
The proof of this lemmas proceeds by applying functional induction on \bap{step}{s}{a}. Then, in \lemref{lemma:soundvalid}, the proof continues by
applying the corresponding subproof of soundness of the function that implements each action; whereas in \lemref{lemma:sounderror}, a subproof about the
existence of a proper error code is provided.

\subsection{Reasoning over the certified reference validation mechanism}
\label{sec:secpropcert}

In this section we present several security properties we have stated and proved about the function $trace$ defined in Section~\ref{sec:excspec:traces}. 

The first property states that in Android \texttt{10}, if an application that is considered to be old (as we defined in
Section~\ref{sec:background:changes:10}) is able to run, then it has been verified and validated by the user previously.

\begin{prop}[Old applications must be verified]
\label{impproperty:oldapp}
 \mbox{} \\
 \footnotesize
$	\forall
	(initState,lastState:\AndroidState)
	(app:AppId)
	(l: list\ \Action)$, \\
$	valid\_state(initState) \rightarrow 
	app \in getInstalledApps(initState) \rightarrow \\
	oldAppNotVerified(a, initState) \rightarrow
	canRun(a, lastState) \rightarrow \\
	last(trace(initState,l),initState) = lastState \rightarrow  \\
	\texttt{uninstall}~app \notin l \rightarrow 
	\texttt{verifyOldApp}~app \in l $ \\ \\
	\textit{The only way for an old application to be able to execute is if the user verified it.}
\end{prop}

The next property establishes that for an application to have \textbf{any} dangerous permission (grouped or ungrouped) it must be explicitly granted to it,
either by the user or automatically by the system.


\begin{prop}[Dangerous permissions must be explicitly granted]
\label{impproperty2}
 \mbox{} \\
 \footnotesize
$	\forall
	(initState,lastState:\AndroidState)
	(app:AppId)
	(p:\Perm)
	(l: list\ \Action)$, \\
$	valid\_state(initState) \rightarrow 
	app \in getInstalledApps(initState) \rightarrow$ \\
$	getPermissionLevel(p)= dangerous \rightarrow 
	permissionIsGrouped(p) = None \rightarrow $ \\
$	appHasPermission(app,p,lastState) \rightarrow$\\
$	\neg appHasPermission(app,p,initState)\rightarrow
	\texttt{uninstall}~app \notin l \rightarrow $ \\
$	last(trace(initState,l),initState) = lastState \rightarrow 
	(\texttt{grant}~p~app \in l~\lor~$ \\
$	\texttt{grantAuto}~p~app \in l)$ \\ \\
	\textit{The only way for an application to get a permission is if the user authorized it, or if the user authorized a group and the system is able to
	automatically grant it}.
\end{prop}

%
The following property formally states that if an application used to have a permission that was later revoked, only re-granting it will allow the application
to have it again.
%


\begin{prop}[Revoked permissions must be regranted]
 \mbox{} \\
 \footnotesize
$	\forall
	(initState,sndState,lastState:\AndroidState)
	(app:\AppId)
	(p:\Perm)
	(l: list~\Action)$, \\
$	valid\_state(initState) \rightarrow
	getPermissionLevel(p) = dangerous \rightarrow$ \\
$   p \notin getDefPermsForApp(app,initState) \rightarrow$ \\
$	step(initState,\texttt{revoke}~p~app).st = sndState \rightarrow$ \\
$	step(initState, \texttt{revoke}~p~app)).resp=ok \rightarrow \\
	\texttt{uninstall}~app \notin l \rightarrow
	\texttt{grant}~p~app \notin l \rightarrow 
	\texttt{grantAuto}~p~app \notin l \rightarrow $\\
$	last(trace(sndState,l),sndState) = lastState \rightarrow$\\
$	\neg appHasPermission(app,p,lastState)$ \\ \\
\textit{ If a permission is revoked from an application, only regranting it will allow the application to have it again.}
\end{prop}




Whenever an application $app$ receives a \texttt{READ/WRITE} permission $perm$, it also receives the right to delegate this
permission to another application, say $app'$, to access that same resource on its behalf. However, if $perm$ is later revoked
from application $app$, there's a chance that $app'$ still has access to that resource, since delegated permissions \textbf{are
not recursively revoked}. The following property formalizes this situation and is a proof that the current specification allows
a behavior which is arguably against the user's will. 

\begin{prop} [Delegated permissions are not recursively revoked] \label{impproperty1} 
 \mbox{} \\
 \footnotesize
$	\forall 
	(s:\AndroidState)
	(p:\Perm)
	(app,app':\AppId)
	(ic,ic': \iComp) 
	(c,c':\Comp)
	(u:\Uri)
	(cp:CProvider), $ \\
$	valid\_state(s) \rightarrow 
	step(s,\texttt{grant}~p~app).resp = ok \rightarrow $ \\
$	getAppFromCmp(c,s) = app \rightarrow 
	getAppFromCmp(c',s) = app' \rightarrow$ \\
$	(ic, c) \in getRunningComponents(s) \rightarrow
	(ic', c') \in getRunningComponents(s) \rightarrow$ \\
$	canGrant(cp,u,s) \rightarrow 
existsRes(cp,u,s) \rightarrow 
componentIsExported(cp) \rightarrow$ \\
$	permissionRequiredRead(cp) = Some~p \rightarrow$ \\
$	\mathsf{let}~opsResult := trace(s,\texttt{[}\texttt{grant}~p~app, \texttt{grantP}~ic~cp~app'~u~Read$,\\ 
$ \texttt{revoke}~p~app\texttt{]}~\mathsf{in}$ 
$	step(last(opsResult,s), \texttt{read}~ic'~cp~u).resp=ok$ \\ \\
\textit{In Android 6 if a permission $p$ is revoked for an application $app$ not necessarily shall it be revoked for the applications to which $app$ delegated $p$.} 
\end{prop}



The purpose of this last property is to show that with runtime permissions introduced after \texttt{Android 6}, certain
assertions on which a developer could rely in previous versions do not hold. For example, a running component may have the right
of starting another one on a certain state, but may not be able to do so at a later time, even though no involved application
was uninstalled. The property still holds on the latest version of Android.

%
\begin{prop}[The right to start an external component is revocable]
 \mbox{} \\
 \footnotesize
$	\forall
	(initState:\AndroidState)
	(l: list \Action)
	(app,app': \AppId)
	(c:\Comp) 
	(act:\Activity) 
	(p:\Perm)$, \\
$	valid\_state(initState) \rightarrow \\
	getPermissionLevel(p) = dangerous \rightarrow$ 
$	permissionIsGrouped(p) = None \rightarrow \\
	app \neq app' \rightarrow $
$	p \notin getDefPermsApp(app,initState) \rightarrow 
	inApp(c,app,initState) \rightarrow$ \\
$	inApp(act,app',initState) \rightarrow 
	cmpProtectedByPerm(act) = Some~p \rightarrow$ \\
$	canStart(c,act,initState) \rightarrow 
	\exists (l: list\ \Action),
	\texttt{uninstall}~app \notin l\ \land$ \\
$	\texttt{uninstall}~app' \notin l \land
	\neg canStart(c,act,last(trace(initState,l), initState))$ \\ \\
\textit{A running component may have the right of starting another one on a certain state, but may not be able to do so at a later time.}
\end{prop}

\section{Related work}
\label{sec:relwork}

    Several analyses have been carried out concerning the security of the Android permission system. Plenty of
    them~\cite{tool:mperm,tool:droidtector,tool:droidsafe,tool:amandroid,tool:ic3,tool:covert} implement a static analysis tool that is capable of detecting
    overprivileges and unwanted information flow on a set of applications. This pragmatic approach may be helpful for Android users to keep their private
    information secure, but no properties about the system can be established. Recently, Mayrhofer \textit{et al.} \cite{DBLP:journals/corr/abs-1904-05572}
    described the Android security platform and documented the complex threat model and ecosystem it needs to operate, but no formal analysis was performed in
    it.

    Few works study the aspects of the permission enforcing framework in a formal way. In particular, Shin \textit{et
        al.}~\cite{Android_formalSecurity2010,DBLP:conf/csreaSAM/ShinKFT10} developed using \texttt{Coq} a framework that represents the Android permission
    system, similarly to what we did. Although, that work does not consider the different types of components, the interaction between a running instance
    and the system, the R/W operation on a content provider, the semantics of the permission delegation mechanism. Also, their work is based on an older
    version of the platform and some novel aspects, like the management of runtime permissions or the verification of legacy applications, are not included.
    Similarly, Bagheri \textit{et al.} \cite{bagheri:alloy} formalized Android permission protocol using \texttt{Alloy}~\cite{Alloy}. The analyses performed,
    however, was based on the ability to automatically find counterexamples provided by the Alloy framework, which the authors claim to be tremendously
    helpful for identifying vulnerabilities. A \texttt{Coq}-based approach like ours, requires more human effort to identify a flaw but provides stronger guarantees
    on security and safety properties.
    Another formal work on Android is CrashSafe~\cite{CrashSafe}, where the authors formalized in \texttt{Coq} the inter-component communication mechanism and proved its
    safety with regard to failures (or \textit{crashes}). This work, similarly to ours, focus on safety properties rather than security ones.

On the other hand, many works have addressed the problem of relating inductively defined relations and executable functions. In particular, Tollitte \emph{et al.}~\cite{Tollitte:2012:CPP} show how to extract a functional implementation from an
inductive specification in \texttt{Coq}, and \cite{Berghofer:2009:TPHOLS} exhibits a similar approach for \texttt{Isabelle}.  
Earlier, alternative approaches such as~\cite{Balaa:2000:TPHOLS,Barthe:2006:FLOPS} aim to provide reasoning principles for executable specifications. 
In \cite{DBLP:journals/scp/AngelisFPP14}, the verification of properties of imperative programs is performed using techniques based on the specialization of constrained logic programs. 
In this work we are able to develop independently the specification of the reference monitor and the implementation of the validation mechanism, considering that \texttt{Coq} provides a reasoning framework based on higher order logic to perform proofs of specifications and programs and a functional programming language.
Other approaches could be considered to develop the formalization. For instance, a logical approach  like the one used in \cite{DBLP:journals/scp/AngelisFPP14}. However, a logical approach does not allow us to have the same functionalities in a unified formal environment.

Specifically, in this work we present a model of a reference monitor and demonstrate properties which shall hold for every correct implementation of the
model. Then, we have developed a functional implementation in  \texttt{Coq} of the reference validation mechanism and proved its
correctness with respect to the specified reference monitor. Applying the program extraction mechanism provided by \texttt{Coq} we have also derived a certified
\texttt{Haskell} prototype of the reference validation mechanism, which can be used to conduct verification activities on actual real implementations of the platform.
    The results presented in this paper extend the ones reported in \cite{LOPSTR:BetarteCGL,DBLP:journals/cleiej/LunaBCSCG18}. We have enriched
    the model presented in \cite{LOPSTR:BetarteCGL,DBLP:journals/cleiej/LunaBCSCG18} so as to consider the changes introduced in Android permission
    system by version \texttt{Nougat}, \texttt{Oreo}, \texttt{Pie} and \texttt{10}.   

\section{Final remarks}
\label{sec:conclusion}

We have enhanced the formal specification considered in our previous work~\cite{LOPSTR:BetarteCGL} with the new features concerning the permission system that
have been added during the later releases of Android. With a conservative approach, we first analyzed the validity of the already formulated properties and then
established new ones about the novel changes; summing up a total of 14 valid properties, without including the auxiliary lemmas that have been separated just
for modularization. Among these properties we included several that aim to highlight how formal methods help to disambiguate unclear behaviours that
may be inferred from an informal specification. For instance, we found a potentially dangerous situation in which
an application can gain access to every dangerous permission that shares group with a normal one, without explicit consent of the user (see
Property~\ref{modelproperty3}). This scenario fits the model (informally) described in the official documentation of the platform.

We also enriched our previous functional implementation of the reference validation mechanism with these new characteristics and updated its correctness proof.
As consequence, the derived \texttt{Haskell} prototype obtained using the program extraction mechanism provided by the proof assistant, has been updated as
well. The full certified code is available in \cite{AndroidCoq:2020}.

One important goal of our work is to keep our formalization up to date with the later versions of Android in order to constitute a reliable framework for
reasoning about its permission system. We aim to help to increase the confidence on Android's security mechanisms by providing certified guarantees about
the enforcement of this measures. The use of idealized models and certified prototypes is a good
step forward but no doubt the definitive step is to be able to provide similar guarantees concerning actual implementations of the platform.
The formal development is about 23k LOC of \texttt{Coq}, including proofs.

On September 8th 2020, Android \texttt{11} was released. This update includes features that continue increasing the security of the device, such as
auto-resetting permissions from unused applications or one-time permissions for the most sensitive resources, like the microphone or camera. A one time
permission is basically a permission as we described in Section~\ref{sec:background} but it only holds for a short period of time, determined by the behaviour
of the app and the user. In future work, we intend to add this features to our model.

\newpage
\bibliography{biblio}

\newcommand{\acceso}{Last access: Oct. 2020}
\begin{thebibliography}{10}

\bibitem{Anderson:1972}
J.~P. Anderson.
\newblock {Computer Security technology planning study}.
\newblock Technical report, Deputy for Command and Management System, USA,
  1972.

\bibitem{fundamentals}
{Android Developers}.
\newblock \textit{Application Fundamentals}.
\newblock Available at:
  \url{http://developer.android.com/guide/components/fundamentals.html}.
\newblock \acceso.

\bibitem{permissions}
{Android Developers}.
\newblock \textit{Permissions}.
\newblock Available at:
  \url{http://developer.android.com/guide/topics/security/permissions.html}.
\newblock \acceso.

\bibitem{protectionLevel}
{Android Developers}.
\newblock \textit{Protection levels}.
\newblock Available at:
  \url{https://developer.android.com/guide/topics/permissions/}.
\newblock \acceso.

\bibitem{tool:covert}
H.~{Bagheri}, A.~{Sadeghi}, J.~{Garcia}, and S.~{Malek}.
\newblock Covert: Compositional analysis of android inter-app permission
  leakage.
\newblock {\em IEEE Transactions on Software Engineering}, 41(9):866--886, Sep.
  2015.

\bibitem{bagheri:alloy}
Hamid Bagheri, Eunsuk Kang, Sam Malek, and Daniel Jackson.
\newblock A formal approach for detection of security flaws in the android
  permission system.
\newblock {\em Formal Aspects of Computing}, 30, 11 2017.

\bibitem{Balaa:2000:TPHOLS}
A.~Balaa and Y.~Bertot.
\newblock Fix-point equations for well-founded recursion in type theory.
\newblock In M.~Aagaard and J.~Harrison, editors, {\em TPHOLs}, volume 1869 of
  {\em LNCS}, pages 1--16. Springer, 2000.

\bibitem{Barthe:2006:FLOPS}
G.~Barthe, J.~Forest, D.~Pichardie, and V.~Rusu.
\newblock Defining and reasoning about recursive functions: A practical tool
  for the coq proof assistant.
\newblock In M.~Hagiya and P.~Wadler, editors, {\em FLOPS}, volume 3945 of {\em
  LNCS}, pages 114--129. Springer, 2006.

\bibitem{Berghofer:2009:TPHOLS}
Stefan Berghofer, Lukas Bulwahn, and Florian Haftmann.
\newblock Turning inductive into equational specifications.
\newblock In S.~Berghofer, T.~Nipkow, C.~Urban, and M.~Wenzel, editors, {\em
  TPHOLs}, volume 5674 of {\em LNCS}, pages 131--146. Springer, 2009.

\bibitem{LOPSTR:BetarteCGL}
Gustavo Betarte, Juan Campo, Felipe Gorostiaga, and Carlos Luna.
\newblock {\em A Certified Reference Validation Mechanism for the Permission
  Model of Android: 27th International Symposium, LOPSTR 2017, Namur, Belgium,
  October 10-12, 2017, Revised Selected Papers}.
\newblock 07 2018.

\bibitem{tool:mperm}
P.~{Chester}, C.~{Jones}, M.~{Wiem Mkaouer}, and D.~E. {Krutz}.
\newblock M-perm: A lightweight detector for android permission gaps.
\newblock In {\em 2017 IEEE/ACM 4th International Conference on Mobile Software
  Engineering and Systems (MOBILESoft)}, pages 217--218, May 2017.

\bibitem{DBLP:journals/scp/AngelisFPP14}
E.~{De Angelis}, F.~Fioravanti, A.~Pettorossi, and M.~Proietti.
\newblock Program verification via iterated specialization.
\newblock {\em Sci. Comput. Program.}, 95:149--175, 2014.

\bibitem{tool:droidsafe}
Michael Gordon, Kim deokhwan, Jeff Perkins, Limei Gilham, Nguyen Nguyen, and
  Martin Rinard.
\newblock Information-flow analysis of android applications in droidsafe.
\newblock 01 2015.

\bibitem{reporteIDC}
International Data~Corporation (IDC).
\newblock Smartphone market share.
\newblock Technical report, International Data Corporation (IDC), 2020.

\bibitem{Alloy}
Daniel Jackson.
\newblock {\em Software Abstractions: Logic, Language, and Analysis}.
\newblock The MIT Press, 2012.

\bibitem{CrashSafe}
Wilayat Khan, Habib Ullah, Aakash Ahmad, Khalid Sultan, Abdullah Alzahrani,
  Sultan Khan, Mohammad Alhumaid, and Sultan Abdulaziz.
\newblock Crashsafe: a formal model for proving crash-safety of android
  applications.
\newblock {\em Human-centric Computing and Information Sciences}, 8, 12 2018.

\bibitem{letouzey04}
P.~Letouzey.
\newblock {\em Programmation fonctionnelle certifi{\'e}e -- L'extraction de
  programmes dans l'assistant {Coq}}.
\newblock PhD thesis, Universit{\'e} Paris-Sud, July 2004.

\bibitem{conf/types/Letouzey02}
Pierre Letouzey.
\newblock {A New Extraction for Coq}.
\newblock In {\em Proceedings of TYPES'02}, volume 2646 of {\em LNCS}, 2003.

\bibitem{AndroidCoq:2020}
Guido~De Luca and Carlos Luna.
\newblock {Formal verification of the security model of Android 10: Coq code}.
\newblock Available at: \url{https://www.fing.edu.uy/~cluna/Android10-Coq.zip}.
\newblock \acceso.

\bibitem{DBLP:journals/cleiej/LunaBCSCG18}
Carlos Luna, Gustavo Betarte, Juan~Diego Campo, Camila Sanz, Maximiliano
  Cristi{\'{a}}, and Felipe Gorostiaga.
\newblock A formal approach for the verification of the permission-based
  security model of android.
\newblock {\em {CLEI} Electron. J.}, 21(2), 2018.

\bibitem{DBLP:journals/corr/abs-1904-05572}
Ren{\'{e}} Mayrhofer, Jeffrey~Vander Stoep, Chad Brubaker, and Nick Kralevich.
\newblock The android platform security model.
\newblock {\em CoRR}, abs/1904.05572, 2019.

\bibitem{tool:ic3}
Damien Octeau, Daniel Luchaup, Matthew Dering, Somesh Jha, and Patrick
  McDaniel.
\newblock Composite constant propagation: Application to android
  inter-component communication analysis.
\newblock In {\em Proceedings of the 37th International Conference on Software
  Engineering - Volume 1}, ICSE '15, pages 77--88, Piscataway, NJ, USA, 2015.
  IEEE Press.

\bibitem{AndroidProy}
{Open Handset Alliance}.
\newblock \textit{Android project}.
\newblock Available at: \url{//source.android.com/}.
\newblock \acceso.

\bibitem{Android_formalSecurity2010}
W.~Shin, S.~Kiyomoto, K.~Fukushima, and T.~Tanaka.
\newblock A formal model to analyze the permission authorization and
  enforcement in the android framework.
\newblock In {\em SocialCom'10}, pages 944--951, Washington, DC, USA, 2010.
  IEEE Computer Society.

\bibitem{DBLP:conf/csreaSAM/ShinKFT10}
W.~Shin, S.~Kiyomoto, K.~Fukushima, and T.~Tanaka.
\newblock A frst step towards automated permission-enforcement analysis of the
  android framework.
\newblock In {\em {SAM} 2010}, pages 323--329. {CSREA} Press, 2010.

\bibitem{coq-manual}
{The Coq Team}.
\newblock {\em The {Coq} Proof Assistant Reference Manual -- Version V8.12.0},
  2020.

\bibitem{Tollitte:2012:CPP}
P.-N. Tollitte, D.~Delahaye, and C.~Dubois.
\newblock Producing certified functional code from inductive specifications.
\newblock In Chris Hawblitzel and Dale Miller, editors, {\em CPP}, volume 7679
  of {\em LNCS}, pages 76--91. Springer, 2012.

\bibitem{tool:amandroid}
Fengguo Wei, Sankardas Roy, Xinming Ou, and Robby.
\newblock Amandroid: A precise and general inter-component data flow analysis
  framework for security vetting of android apps.
\newblock {\em ACM Transactions on Privacy and Security}, 21:1--32, 04 2018.

\bibitem{tool:droidtector}
S.~{Wu} and J.~{Liu}.
\newblock Overprivileged permission detection for android applications.
\newblock In {\em ICC 2019 - 2019 IEEE International Conference on
  Communications (ICC)}, pages 1--6, May 2019.

\end{thebibliography}

\end{document}